\documentclass[fleqn,usenatbib]{mnras}

\usepackage{newtxtext,newtxmath}

\usepackage[T1]{fontenc}
\usepackage{ae,aecompl}
\usepackage{graphicx}
\usepackage{txfonts}
\usepackage{natbib}
\usepackage{enumitem}
\usepackage{rotating}
\usepackage{pdflscape}

\bibpunct{(}{)}{;}{a}{}{,}

 \newcommand{\mic}{$\mu$m}

\usepackage{graphicx}	
\usepackage{amsmath}	
\usepackage{amssymb}	

\title[ALMA confirmation of $z \sim 3-5$ passive candidates]{Passive galaxies in the early Universe: ALMA confirmation of $z \sim 3-5$ candidates in the CANDELS GOODS-South field}

\author[P. Santini et al.]{
P. Santini$^{1}$\thanks{E-mail: paola.santini@inaf.it},
E. Merlin$^{1}$,
A. Fontana$^{1}$,
B. Magnelli$^{2}$,
D. Paris$^{1}$,
M. Castellano$^{1}$,
\newauthor A. Grazian$^{1}$,
L. Pentericci$^{1}$,
S. Pilo$^{1}$,
and M. Torelli$^{1}$
\\
$^{1}$INAF - Osservatorio Astronomico di Roma, via di Frascati 33, I-00078 Monte Porzio Catone (Roma), Italy\\
$^{2}$Argelander-Institut f\"{u}r Astronomie, Universit\"{a}t Bonn, Auf dem H\"{u}gel 71, D-53121 Bonn, Germany
}

\date{Accepted XXX. Received YYY; in original form ZZZ}

\pubyear{2018}

\begin{document}
\label{firstpage}
\pagerange{\pageref{firstpage}--\pageref{lastpage}}
\maketitle


\begin{abstract}
  The selection of red, passive galaxies in the early Universe is very
  challenging, especially beyond $z \sim 3$, and it is crucial to
  constrain theoretical modelling of the processes responsible for
  their rapid assembly and abrupt shut-down of the star formation.  We
  present here the analysis of ALMA archival observations of 26 out of
  the 30 galaxies in the deep CANDELS GOODS-South field that we
  identified as passive at $z\sim3-5$ by means of a careful and
  conservative SED fitting analysis. ALMA data are used to verify the
  potential contamination from red, dusty but star--forming sources
  that could enter the sample due to similar optical--nearIR colours.
  With the exception of a few marginal detections at <3$\sigma$, we
  could only infer upper limits, both on individual sources and on the
  stacks.  We translated the ALMA continuum measurements into
  corresponding SFRs, using a variety of far-IR models.  These SFRs
  are compared with those predicted by secondary star-forming
  solutions of the optical fits and with the expected position of the
  star formation Main Sequence.  This analysis confirms the passive
  nature of 9 candidates with high confidence and suggests that the
  classification is correct for at least half of the sample in a
  statistical sense. For the remaining sources the analysis remain
  inconclusive because available ALMA data is not deep enough,
  although the stacking results corroborate their passive nature.
  Despite the uncertainties, this work provides decisive support to
  the existence of passive galaxies beyond $z \sim 3$. 
 \end{abstract}



\section{Introduction}\label{sec:intro}

The existence of massive, passively evolving galaxies at high redshift
represents an arduous challenge to theoretical models of galaxy
formation, that struggle to reproduce the observations
\citep{fontana09,vogelsberger14,feldmann16}. The abundance of these
galaxies at different epochs is a crucial observable to constrain the
different physical processes responsible for their rapid assembly and
for the abrupt shut-down of their star formation activity. Theoretical
simulations are indeed very sensitive to the detailed modeling of
processes such as merger-driven starbursts or feedback
\citep[e.g.][]{menci06,hopkins08,choi15}.

To better understand these delicate physical processes, it is
important to define reliable samples to which compare theoretical
predictions. This is not a straightforward task, especially at high
redshift, where highly dust-enshrouded galaxies are much more abundant
and well mimic the emission of red evolved ones \citep{brammer09}. A
number of selection criteria have been developed to this aim, such as
colour-colour diagrams
\citep[e.g.][]{franx03,daddi04,wuyts07,martis16} or SED fitting
\citep[e.g.][]{grazian07,fontana09}, and are mostly limited to
relatively low-intermediate redshifts.  Some of the candidates have
been spectroscopically confirmed
\citep[e.g.][]{cimatti04,onodera12,whitaker13}. In particular, the
recent detection of a quiescent galaxy at $z\simeq 3.7$ provided
crucial evidence of the existence of such objects even at $z>3$
\citep{glazebrook17,schreiber18b}.

While being relatively easy to implement, especially at high redshift
where other techniques become hard and sometimes not applicable,
selections based on colours may suffer from incompleteness. This is
clearly demonstrated by our previous work \citep[][M18
hereafter]{merlin18}, where we show that galaxies that have undergone
an abrupt truncation of their star formation activity can remain
outside the passive selection region of the $U-V$ vs $V-J$ diagram
\citep{williams09} for a few hundreds Myr. In M18 we performed a very
accurate and conservative selection based on SED fitting, with an
appropriate choice of the model star formation history (SFH), and
selected a sample of 30 passive, red and dead galaxies at $z>3$ in the
GOODS-South field. Our analysis showed that the reliability of the
selection depends also crucially on the details of the SED fitting
method, such as the inclusion of emission lines or the full inclusion
of redshift uncertainties, that may decrease the size of the sample by
large factors. For this reason, and taking into account also other
possible degeneracies between red dusty/star--forming and passive
solutions, a more stringent verdict upon the lack of star formation is
achievable through far-IR/submm observations, which are able to sample
the cold dust emission, expected to be prominent in star--forming
galaxies. In M18 we performed a sanity check by means of {\it
  Herschel} data, and found detection for 2 out of 30 candidates,
potentially caused by hot dust emission from an AGN hosted in these
two galaxies. However, {\it Herschel} observations only allow the
detection of few, extremely star--forming galaxies at these redshifts,
while normal, Main Sequence (MS hereafter) galaxies would remain
undetected in any case.  In this paper, we make use of the rich ALMA
archive to search for cold dust emission, hinting on-going star
formation, around our candidates, and exploit the sub-mm inferred
(limits on the) SFR to validate our classification, both on an
individual basis and in a statistical sense.

The paper is organized as follows. We summarize our previous work and
candidate selection in Sect.~\ref{sec:targetsel}, describe the ALMA
observations in Sect.~\ref{sec:alma}, and derive the expected SFR
based on these observations in Sect.~\ref{sec:sfr}. Finally, we
present our results in Sect.~\ref{sec:results} and draw our
conclusions in Sect.~\ref{sec:concl}. In the following, we adopt the
$\Lambda$-CDM concordance cosmological model (H$_0$=70 km/s/Mpc,
$\Omega_M$=$0.3$ and $\Omega_{\Lambda}$=0.7) and a \cite{salpeter55}
IMF.  All magnitudes are in the AB system.

\section{Candidates selection } \label{sec:targetsel}

We briefly summarize here the strategy pursued in M18 to select
candidate passive galaxies in the CANDELS GOODS-S field by means of
SED fitting.

Our selection takes advantage of the deep and high quality photometry
available in this field as well as of the sophisticated photometric
measure techniques adopted \citep{merlin15,merlin16}. In M18 we
demonstrated that the known criterion based on the rest frame $U-V$ vs
$V-J$ colours ($UVJ$ in the following) suffers from uncertainties due
to the high redshift and extremely red colours of the desired
candidates. In addition, we showed that the $UVJ$ criterion is
physically inappropriate to take into account the short timescales for
galaxies to become quiescent at $z>3$, especially using the standard
exponentially declining laws to model star--formation histories.

To face these issues, we adopted a SED-fitting technique assuming a
``top-hat'' star formation history, characterized by a period of
constant star formation followed by an abrupt truncation of the star
formation, that is set to zero thereafter. To estimate the reliability
of our candidates we adopted a full statistical analysis, implementing
a strict criterion based on the probability $P$ of the $\chi^2$
resulting from the fitted solution. For a galaxy to be selected as
passive we require that the best fit solution is passive and has a
probability $P$($\chi^2_{Q}$)>30\% \textit{and} that no star-forming
solution with $P$($\chi^2_{SF}$)>5\% exists.

We have implemented this approach in three different flavours.  First
we have adopted naked \cite{bc03} models without any inclusion of
nebular emission, and fixing the redshift at the photometric one
\citep[e.g.,][]{grazian07,fontana09}. This choice results in a sample
of 30 objects (dubbed $S_0$ hereafter). We have then added emission
lines, self-computed on the basis of the ionizing flux of each
template, as described in \cite{castellano14} and
\cite{schaerer09}. The inclusion of solutions with strong emission
lines changes the predicted shape of the spectral slope in the reddest
bands, and strongly decreases the number of candidates to 10 ($S_1$
sample).  Finally, we have also let the redshift free to vary, and
removed from the sample the objects that have a plausible
star--forming solution at a different redshift. This way we are left
with only 2 objects in the sample ($S_2$ sample).

This drastic reduction of the number of the 'bona-fide' candidates as
we vary the spectral models used for the star--forming templates is
found also in the whole CANDELS field, as we present in a companion
paper (Merlin et al. 2018, in prep).  We remark that our method is
quite conservative, as it retains only the objects that have both a
quiescent best-fit solution and no plausible star--forming solutions.

Far-infrared data are fundamental to exclude the potential
star--forming solutions for our candidates. In M18 we have searched
for {\it Herschel} counterparts of our candidates and found detections
for two of them. As discussed in Sect.~\ref{sec:results}, they turned
out to be potentially obscured AGNs rather than star--forming
galaxies.  {\it Herschel} images are, however, not deep enough to
probe normal star--forming galaxies at $z>3$.  Deep ALMA observations
therefore provide, at present, the only tool for validating the
passive solutions for our candidates.  For this reason we have
searched the ALMA archive for observations of the whole $S_0$ sample.

\section{ALMA observations and data  analysis}\label{sec:alma}

\begin{table}
\centering
\caption{List of ALMA programs used in this analysis and corresponding PI, ALMA band and resolution.}
\begin{tabular} {cccc}
\hline 
\hline 
\noalign{\smallskip} 
ALMA program & PI & Band & Beam \\ 
 & &  &  [arcsec x arcsec]\\ 
\noalign{\smallskip} 
\hline 
\noalign{\smallskip} 
2012.1.00173.S  &  J. Dunlop  &  6  & 0.62x0.52 \\
2012.1.00869.S  &  J. Mullaney  &  7  & 0.74x0.62 \\
2013.1.00718.S  &  M. Aravena  &  6  & 1.44x0.73 \\
2013.1.01292.S  &  R. Leiton  &  7  & 0.61x0.58 \\
2015.1.00098.S  &  K. Kohno  &  6  & 0.60x0.60 \\
2015.1.00543.S  &  D. Elbaz  &  6  & 0.60x0.60 \\
2015.1.00664.S  &  K. Tadaki  &  6  & 0.72x0.58 \\
2015.1.00870.S  &  T. Wiklind  &  7  & 0.70x0.61 \\
2015.1.01074.S  &  H. Inami  &  7  & 0.67x0.59 \\
2015.1.01495.S  &  T. Wang  &  7  & 0.63x0.58 \\
\noalign{\smallskip}
\hline 
\end{tabular}
\label{tab:list_programs}
\end{table}

We have inspected the ALMA archive at the positions of our candidates,
and found public observations for 26 out of 30 sources, belonging to a
number of different programs, listed in Table
~\ref{tab:list_programs}.

Among the 26 sources observed by ALMA are one of the two objects
belonging to the $S_2$ sample (ID10578) and further 7 belonging to the
$S_1$ sample ( ID2782, ID3912, ID8785, ID9209, ID17749, ID18180 and
ID23626). The observed sources are listed in Table~\ref{tab:list}.

The observations have been carried out in Band 6 and Band 7, with
different sensitivities, setups and configurations.  We have not used
Band 3 and Band 4 observations as they do not add further information
to the analysis (due to their shallowness and/or their dearth).  Most
of the sources are covered by more than one program, either in the
same band or in a different one. We have stacked the sources observed
more than once in the same band and we have combined the results from
the different bands, as explained below.

\begin{figure}
\includegraphics[width=\columnwidth,viewport=1 80 710 500]{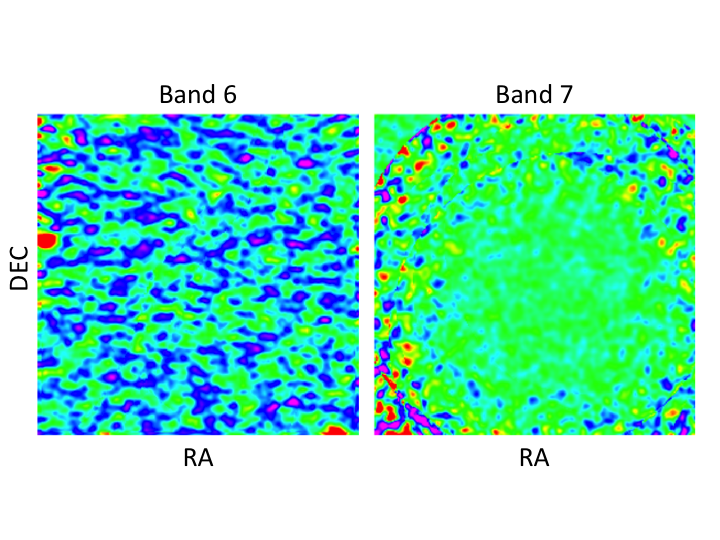}
        \caption{Stacked continuum images (25\arcsec $\times$ 25\arcsec) in Band 6 ({\it left)} and Band 7 ({\it right)}, of 21
          and 16 sources, respectively, at $3<z<5$.}
    \label{fig:stacks}
\end{figure}

Observations were calibrated with CASA \citep{mcmullin07} using
scripts provided by the ALMA project.  Imaging was then performed
using the multi-frequency synthesis algorithm implemented within the
CASA task \textit{CLEAN}. We used "natural" weightings and
\textit{uv}-tapers when needed, producing images with spatial
resolution $>0.6$\arcsec (see Table~\ref{tab:list_programs}). This
resolution was chosen following results from \cite{franco18}, who
demonstrated that all sources are unresolved in their GOODS-S
0.6\arcsec-tapered mosaic (from which most of our sources are taken)
and that the average size of galaxies at submillimetre wavelengths is
$0.3\pm0.1$\arcsec.  Our galaxies being in addition relatively compact
in the near-IR (average $R$\textsubscript{1/2}$\sim$0.22\arcsec in the
$H$ band), we could securely assume that all were point-like sources
in our ALMA images.

With this point-like source assumption, the flux of each candidate
corresponds simply to the pixel value at its position in the primary
beam-corrected ALMA image (in unit of mJy/beam). The associated flux
error was measured by taking the standard deviation of all pixels in
the map with similar coverage, i.e., pixels corresponding to primary
beam corrections within +/-10\% to that at the position of the
candidate. In this procedure, we excluded pixels within one FWHM of
the candidate and we applied a 3$\sigma$ clipping method to avoid
biases from the candidate and any other sources.

We report in Table~\ref{tab:list} the flux measured for each
object. As can be seen, we found no high-confidence (i.e.,
$>$3$\sigma$) detection on ALMA images for any of the candidates.  For
sources observed more than once in the same band, we have stacked the
inferred flux densities by averaging the fluxes measured from
different programs weighting them with the associated errors (the
final sensitivity per beam was inferred as the standard error on the
weighted mean).  Despite the improved sensitivity achieved by
stacking, none of the sources is detected at a significant confidence
level, with only two sources marginally detected at
$\sim 2-2.5 \sigma$ in Band 7. With the exception of additional 5
measurements barely above the noise level (1-1.5$\sigma$), for the
rest we could only infer upper limits.

We note that the number of 1 and 2$\sigma$ detections is consistent
with a normal distribution of the signal-to-noise ratio, i.e., is
consistent with a sample of undetected sources: indeed, out of our 37
measurements (26 sources, some of which observed in both bands), one
would expect $\sim 6$ sources in the upper ($>1\sigma$) tail of the
S/N distribution, of which slightly less than one at $>2\sigma$.

Finally we stacked all sources observed in the same band both over the
entire redshift range and divided in two redshift bins ($3<z<4$ and
$4<z<5$). No detection is obtained in Band 7.  Sources in Band 6 are
only marginally detected ($\lesssim 2\sigma$, in both the $3<z<4$ and
$3<z<5$ redshift bins). When stacking only sources which are
individually undetected, a flux comparable with the noise level is
measured at $3<z<5$ in Band 6, while no detection is obtained at
$3<z<4$. The fact that no detection emerges even from the stacks
allows us to exclude the possibility that a significant fraction of
the undetected objects has flux at >1$\sigma$.

We show in Fig.~\ref{fig:stacks} the stacked images in Band 6 and 7
obtained by averaging all individual images weighting them with the
corresponding rms.  The lack of detection at center can be clearly
seen.

Table~\ref{tab:list} lists the ALMA program, band, actual measured
flux and image sensitivity per beam 
for each of the 26 sources, together with their best-fit redshift and
stellar mass.  We also list the stacking results for the sources
observed in the two ALMA bands, both considering all sources and only
the undetected ones.

\section{ALMA predictions on the SFR} \label{sec:sfr}

\begin{figure*}
	\includegraphics[width=2.0\columnwidth,clip,viewport=1 1 720 320]{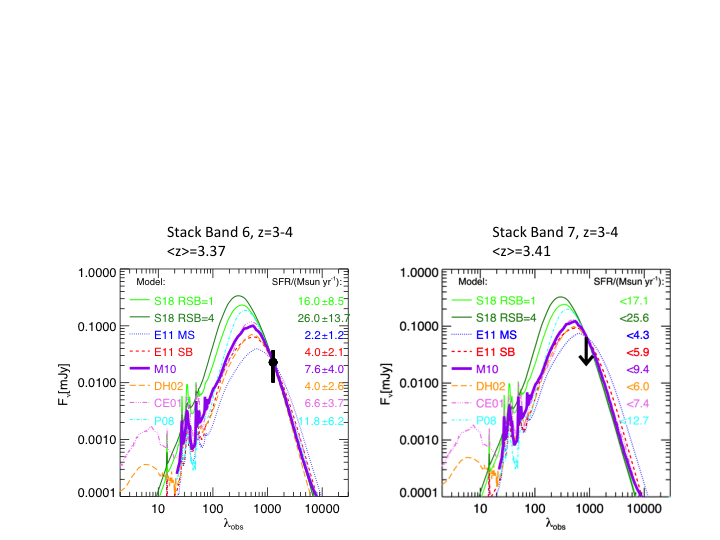}
        \caption{Best fit to the ALMA (limiting) flux measured by
          stacking all $3<z<4$ sources in Band 6 ($left$) and Band 7
          ($right$). Our reference model (M10) is shown by the thick
          purple lines, while the others are represented by thin lines
          colour-coded and characterized by different linestyles
          according to the legend on the left. On the right hand side
          of each panel, we report the corresponding SFR.}
    \label{fig:bestfits}
\end{figure*}

To quantify whether these observations are effective in discriminating
against star--forming solutions, we have used ALMA observations to
infer estimates of, or in most cases upper limits on, the
dust-obscured SFR.  To this aim, we computed the $1\sigma$ upper
limits on the flux as the rms for any source with zero or negative
measured flux, and as measured flux plus the rms for any source with a
small amount of positive flux. If the measured flux is larger than the
rms, the measurement is treated as a marginal detection. When using
ALMA observations to individually validate the candidates, we adopt
more conservative $3\sigma$ limits.

For candidates whose flux has been obtained by combining different
ALMA images with slightly different frequency settings within the same
band, we have computed the final band as a weighted average of the
individual spectral windows covered by the different observations.

\begin{table*}
\centering
\caption{Passive galaxy candidates covered by ALMA observations,  ordered by their reliability ($S_2$, $S_1$ and $S_0$ sample - see text- separated by lines). 
  The SFRs are obtained from the M10 template.  1$\sigma$ upper limits are provided for $<1\sigma$ detections. \newline
  Notes:  
  The 'C' next to the ID mark candidates that are individually confirmed with high confidence. 
  Redshift and stellar masses for the stacks have been obtained as a weighted average. The uncertainty on the stacked stellar mass has been computed from the weighted average relative error. 
}  
\begin{tabular} {lcccccccc}
\hline 
\hline 
\noalign{\smallskip} 
 ID & $z$ & Stellar mass  & Band & ALMA program(s) & Measured flux & Sensitivity per beam & SFR \\ 
 & & $\log$ ($M$/$M_\odot$) &  &  & [mJy/beam] & [mJy/beam] & [$M_\odot$/$yr$] \\
\noalign{\smallskip} 
\hline 
\hline 
\noalign{\smallskip} 
 10578 ~C &3.06 & 11.38$_{-0.19}^{+0.16}$ &  6 & 2015.1.00098.S + 2015.1.00543.S &    0.14 &    0.09 & 38.2$\pm$24.2\\
 &  &  &  7 & 2015.1.01074.S  &    0.04 &    0.84 & <130.0\\
 \hline 
  2782 ~~~C & 3.47 & 10.84$_{-0.20}^{+0.09}$ &  7 & 2015.1.00870.S &    0.04 &    0.11 & <22.5\\
  3912 & 4.08 & 10.56$_{-0.25}^{+0.23}$ &  6 & 2015.1.00543.S &   -0.08 &    0.17 & <42.7\\
  8785 & 3.98 & 10.59$_{-0.22}^{+0.16}$ &  6 & 2015.1.00543.S &   -0.39 &    0.22 & <56.4\\
  9209 ~~~C & 4.55 & 10.96$_{-0.83}^{+0.11}$ &  6 & 2015.1.00543.S &   -0.03 &    0.16 & <41.2\\
 &  &  &  7 & 2015.1.01074.S + 2013.1.01292.S &    0.10 &    0.28 & <58.8\\
 17749 ~C & 3.73 & 11.04$_{-0.28}^{+0.11}$ &  6 & 2015.1.00098.S + 2015.1.00543.S &    0.12 &    0.08 & 32.4$\pm$20.4\\
 &  &  &  7 & 2013.1.01292.S &    0.04 &    0.30 & <48.7\\
 18180 ~C & 3.61 & 10.95$_{-0.24}^{+0.10}$ &  6 & 2015.1.00098.S + 2015.1.00543.S &    0.08 &    0.08 & 21.5$\pm$20.3\\
 &  &  &  7 & 2013.1.01292.S &   -0.48 &    0.28 & <40.6\\
 23626 & 4.64 & 10.88$_{-0.18}^{+0.14}$ &  7 & 2013.1.01292.S &    0.16 &    0.30 & <74.0\\
 \hline 
  2608 & 3.58 &  9.65$_{-0.23}^{+0.10}$ &  7 & 2015.1.00870.S &   -0.10 &    0.19 & <27.5\\
  3973 ~~~C & 3.67 & 11.27$_{-0.27}^{+0.04}$ &  7 & 2013.1.01292.S &    0.74 &    0.29 & 108.8$\pm$43.3\\
  4503 ~~~C & 3.52 & 11.15$_{-0.23}^{+0.10}$ &  6 & 2015.1.00543.S &   -0.56 &    0.31 & <77.4\\
 &  &  &  7 & 2013.1.01292.S &   -0.04 &    0.29 & <42.3\\
  4587 & 3.58 &  9.74$_{-0.16}^{+0.25}$ &  6 & 2015.1.00543.S &   -0.08 &    0.17 & <43.6\\
  5592 & 4.45 & 10.48$_{-0.34}^{+0.20}$ &  6 & 2015.1.00543.S + 2015.1.00870.S + 2015.1.00664.S &    0.05 &    0.05 & <29.6\\
  6407 & 4.74 & 10.20$_{-0.11}^{+0.23}$ &  6 & 2015.1.00543.S &   -0.13 &    0.17 & <42.9\\
 &  &  &  7 & 2012.1.00869.S &   -0.20 &    0.37 & <59.2\\
  7526 & 3.42 & 10.56$_{-0.29}^{+0.17}$ &  6 & 2015.1.00543.S &    0.10 &    0.23 & <84.5\\
 &  &  &  7 & 2015.1.00870.S &    0.10 &    0.12 & <31.5\\
  7688 & 3.35 & 10.36$_{-0.31}^{+0.19}$ &  6 & 2015.1.00543.S &    0.25 &    0.22 & 63.1$\pm$56.6\\
 &  &  &  7 & 2015.1.00870.S &    0.09 &    0.14 & <34.3\\
  8242 & 3.18 &  9.82$_{-0.18}^{+0.11}$ &  6 & 2015.1.00543.S &    0.18 &    0.23 & <104.4\\
  9091 & 3.30 &  9.45$_{-0.15}^{+0.29}$ &  6 & 2015.1.00543.S &    0.07 &    0.17 & <62.6\\
 &  &  &  7 & 2012.1.00869.S &   -0.08 &    0.11 & <16.0\\
 10759 & 3.07 &  8.96$_{-0.53}^{+0.36}$ &  6 & 2015.1.00543.S &    0.07 &    0.22 & <75.4\\
 12178 & 3.28 & 10.61$_{-0.13}^{+0.15}$ &  6 & 2015.1.00543.S &    0.13 &    0.17 & <78.1\\
 15457 & 3.41 &  9.64$_{-0.07}^{+0.22}$ &  6 & 2015.1.00543.S + 2015.1.00098.S + 2012.1.00173.S &    0.03 &    0.03 & 11.2$\pm$11.0\\
 &  &  &  7 & 2015.1.01074.S &    1.79 &    0.79 & 265.0$\pm$117.3\\
16506 ~C & 3.34 &  9.70$_{-0.06}^{+0.23}$ &  6 & 2015.1.00543.S + 2015.1.00098.S + 2012.1.00173.S  &    &    & \\
  &  &  &   &  + 2013.1.00718.S &    0.01 &    0.02 & <9.2\\
 19301 & 3.60 & 10.06$_{-0.27}^{+0.21}$ &  6 & 2015.1.00098.S + 2015.1.00543.S &    0.01 &    0.08 & <26.2\\
 19446 & 3.25 & 10.30$_{-0.33}^{+0.07}$ &  6 & 2015.1.00098.S + 2015.1.00543.S &    0.03 &    0.09 & <31.9\\
 19505 ~C & 3.33 & 10.67$_{-0.17}^{+0.05}$ &  6 & 2015.1.00870.S &    0.02 &    0.04 & <20.6\\
 &  &  &  7 & 2015.1.00870.S &   -0.05 &    0.11 & <16.8\\
 22610 & 3.22 &  9.98$_{-0.17}^{+0.17}$ &  7 & 2015.1.01495.S + 2015.1.01074.S &    0.09 &    0.24 & <48.9\\
 \noalign{\smallskip}
 \hline 
 \noalign{\smallskip}
 & 3.44 & 10.31$_{-0.11}^{+0.09}$ &  6 & Stack all Band 6 sources at 3<z<5 &    0.02 &    0.01 & 7.4$\pm$3.8\\
 & 3.37 & 10.29$_{-0.10}^{+0.08}$ &  6 & Stack all Band 6 sources at 3<z<4 &    0.02 &    0.01 & 7.6$\pm$4.0\\
 & 4.45 & 10.54$_{-0.33}^{+0.19}$ &  6 & Stack all Band 6 sources at 4<z<5 &    0.02 &    0.05 & <18.9\\
 & 3.49 & 10.66$_{-0.22}^{+0.15}$ &  7 & Stack all Band 7 sources at 3<z<5 &    0.02 &    0.05 & <9.5\\
 & 3.41 & 10.64$_{-0.21}^{+0.14}$ &  7 & Stack all Band 7 sources at 3<z<4 &    0.02 &    0.05 & <9.4\\
 & 4.63 & 10.84$_{-0.33}^{+0.18}$ &  7 & Stack all Band 7 sources at 4<z<5 &    0.05 &    0.18 & <36.5\\
 & 3.44 & 10.15$_{-0.11}^{+0.08}$ &  6 & Stack undetected Band 6 sources at 3<z<5 &    0.01 &    0.01 & 4.3$\pm$4.3\\
 & 3.35 & 10.09$_{-0.09}^{+0.07}$ &  6 & Stack undetected Band 6 sources at 3<z<4 &    0.01 &    0.01 & <8.7\\
 & 4.45 & 10.54$_{-0.33}^{+0.19}$ &  6 & Stack undetected Band 6 sources at 4<z<5 &    0.02 &    0.05 & <18.9\\
 & 3.48 & 10.63$_{-0.22}^{+0.15}$ &  7 & Stack undetected Band 7 sources at 3<z<5 &   -0.00 &    0.05 & <6.7\\
 & 3.40 & 10.61$_{-0.21}^{+0.14}$ &  7 & Stack undetected Band 7 sources at 3<z<4 &   -0.01 &    0.05 & <6.9\\
 & 4.63 & 10.84$_{-0.33}^{+0.18}$ &  7 & Stack undetected Band 7 sources at 4<z<5 &    0.05 &    0.18 & <36.5\\
\noalign{\smallskip}
\hline 
\end{tabular}
\label{tab:list}
\end{table*}

\begin{table}
\centering
\caption{ALMA based (limiting) SFR obtained by combining the results from observations in different bands, when available. The lower part lists the stacking results when only considering the final (i.e., obtained after combining all data) undetected sources.
}  
\begin{tabular} {ccc}
\hline 
\hline 
\noalign{\smallskip} 
 ID & Combined SFR[$M_\odot$/$yr$] & Band \\
\noalign{\smallskip} 
\hline 
\noalign{\smallskip} 
 10578 & 38.2$\pm$24.2 &      6\\
  2782 & <22.5 &      7\\
  3912 & <42.7 &      6\\
  8785 & <56.4 &      6\\
  9209 & <41.2 &      6\\
 17749 & 32.4$\pm$20.4 &      6\\
 18180 & 21.5$\pm$20.3 &      6\\
 23626 & <74.0 &      7\\
  2608 & <27.5 &      7\\
  3973 & 108.8$\pm$43.3 &      7\\
  4503 & <42.3 &      7\\
  4587 & <43.6 &      6\\
  5592 & <29.6 &      6\\
  6407 & <42.9 &      6\\
  7526 & <31.5 &      7\\
  7688 & 63.1$\pm$56.6 &      6\\
  8242 & <104.4 &      6\\
  9091 & <16.0 &      7\\
 10759 & <75.4 &      6\\
 12178 & <78.1 &      6\\
 15457 & 13.4$\pm$11.0 &     6 and 7\\
 16506 & <9.2 &      6\\
 19301 & <26.2 &      6\\
 19446 & <31.9 &      6\\
 19505 & <16.8 &      7\\
 22610 & <48.9 &      7\\
 \noalign{\smallskip}
 \hline 
 \noalign{\smallskip}
Stack undetected Band 6 & &      \\
 sources at 3<z<4 & <8.7 &      6\\
Stack undetected Band 6 &  &      \\
 sources at 4<z<5 & <18.9 &      6\\
Stack undetected Band 7 &  &      \\
 sources at 3<z<4 & <7.6 &      7\\
\noalign{\smallskip}
\hline 
\end{tabular}
\label{tab:sfr}
\end{table}

The SFR has been computed from the total infrared luminosity between 3
and 1100~\mic~adopting the calibration of \cite{kennicutt12}, adjusted
to a Salpeter IMF using their conversion factor.  We note that we do
not take into account the contribution from old stellar populations to
dust heating, that may be not negligible if our candidates are truly
passive. The true SFRs are therefore likely to be lower than those
inferred by us, hence our results have to be considered as
conservative.

To obtain an estimate of or an upper limit on the total infrared
luminosity we have adopted a number of different models available in
the literature. As reference model we used the average SED of
\cite{michalowski10a} (M10), based on a sample of 76 SMGs at
$0.01<z<3.6$ fitted with the GRASIL \citep{silva98} model. We then
considered the average SMG template of \cite{pope08} (P08), the two
average SEDs fitted by \cite{elbaz11} (E11) for MS and Starburst (SB)
galaxies, and the full libraries of \cite{ce01} (CE01), \cite{dh02}
(DH02) and \cite{schreiber18} (S18). To reduce the number of free
parameters of the latter library, we constrained the dust temperature
and IR8 parameter (=$L_{IR}$/$L_{8\mu m}$) based on the source
redshift following the recipes provided by the authors; we considered
a template for a MS galaxy ($R_{SB}$=$SFR$/$SFR_{MS}$=1) and one for a
starburst galaxy with $R_{SB}$=4.  The results obtained from the M10
model are reported in Table~\ref{tab:list}. As we show later, the
sub-mm based SFRs vary only mildly with redshift. This allows us to
compute the SFR also for our stacked sources at the average redshift
of each group of them, that we report in the lower part of
Table~\ref{tab:list}.

The application of different models yields a range of resulting SFRs,
that are typically in the range $10-50 M_\odot$/$yr$, but that may
span nearly a decade if we consider the full range of adopted models,
especially when the two extreme S18 models are considered. This is
clearly shown in Fig.~\ref{fig:bestfits} where we show the different
IR SEDs obtained after normalizing the models to the observed
(limiting) flux of the average source, obtained by stacking all
candidates at $3<z<4$ in the two ALMA bands. For a given ALMA flux,
the total IR spectrum resulting from each model is clearly different,
and hence the resulting SFR, that is derived from the total IR
luminosity. We note in particular that the two S18 models predict a
significantly higher total IR luminosity because the peak of their SED
is at lower wavelengths than the other models.  The S18 templates,
built on {\it Herschel} and ALMA $0.5<z<4$ detections, assume an
increasing dust temperature with redshift, resulting in a
mass-weighted temperature around 40-50 K at the redshift of our
galaxies (and an even higher expected luminosity-weighted temperature,
as discussed at length in \citealt{scoville16}). We note, however,
that our sources are undetected (or only marginally detected) at
submillimeter wavelengths, hence not representative of the sample
adopted to build the library. Moreover, the recent work of
\cite{gobat18} fits a dust temperature of 21-25 K for $z\sim 1.8$
quiescent galaxies. For these reasons, we believe that the S18 library
is likely inappropriate and overestimates the SFR for our
candidates. Finally, it is also clear from Fig.~\ref{fig:bestfits}
that longer wavelength bands yield the most uncertain total IR
luminosity, as they fall far from the peak of the grey body emission.

In the following we will refer to M10 as our reference model since it
was built on red sub-mm sources (as could be the case of our
candidates, though mostly below the sensitivity of ALMA images) at
high redshift. Moreover, this model predicts SFRs that are somewhat
intermediate with respect to all other models considered.  Using our
reference M10 model we find that the ALMA observations constrain the
SFR of our objects to be typically below 40-50 $M_\odot$/$yr$.

We combined measurements obtained from different bands in order to end
up with a single value of the SFR for each source and for each
model. When available, we used the detections or a weighted average of
them, albeit we remind that these detections are in any case below
3$\sigma$. We verified their consistency with the limits, when limits
and detections are available for the same source.  When only upper
limits are available, we took the most stringent one. The final SFRs
used for the analysis are listed in Table~\ref{tab:sfr}.

\section{Validating the passive solutions} \label{sec:results}

We finally use the SFRs inferred above to validate the quiescent
nature of our candidates, both individually and of the whole sample in
a statistical sense.

\subsection{Validation of robust individual candidates}

We compare the ALMA-based SFRs to the SFRs predicted by the optical
fit. Basically, to test whether our candidates were erroneously
best-fitted by passive templates, we can check whether the alternative
star--forming solutions are compatible or not with ALMA results.
  
\begin{figure*}
	\includegraphics[width=2\columnwidth,clip,viewport=40 50 570 750]{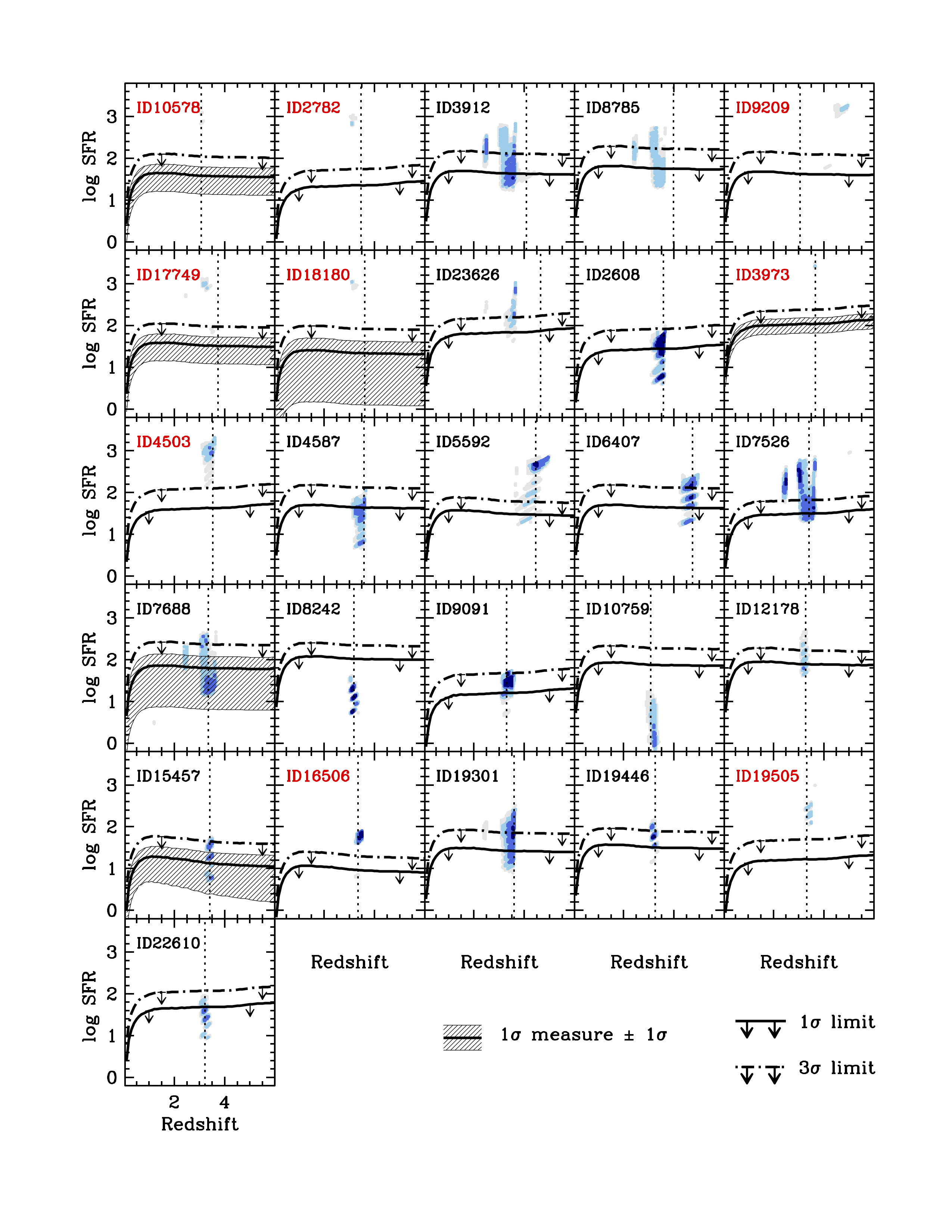}
        \caption{SFR confidence interval or upper limits on the SFR at
          different redshifts based on ALMA observations according to
          our reference model M10.  We show 1$\sigma$ results as solid
          curves and 3$\sigma$ ones as dot-dashed curves.  The
          vertical dotted line indicates the best-fit CANDELS
          photo-z. Blues dots, from lighter to darker shades, show the
          SFR inferred by the optical best-fit by considering
          star--forming solutions at different redshifts (see text),
          associated with probabilities 5-20\%, 20-50\% and $>$50\%,
          respectively. Light gray dots show lower probability
          solutions (P<5\%).  The ID is printed in red colour for
          sources individually confirmed with high confidence.}
    \label{fig:limits}
\end{figure*}

To this aim, we have performed the SFR computation at any redshift
between 0 and 6 to account for possible uncertainties in the photo-z
fitting.  The outcome of the analysis is reported in
Fig.~\ref{fig:limits}, where we show the resulting (limiting) SFR at
all possible redshifts for all 26 sources.  As mentioned above, it is
clear that the resulting SFR is an almost flat function of redshift
(unsurprisingly, as the negative K-correction at sub-mm and mm
wavelengths almost compensates for cosmological dimming).  We compare
the ALMA-based SFRs (at any possible redshift) to the SED fitting SFRs
predicted by the star--forming solutions, ranked with their
probability of reproducing the observed SED, as measured by the
probability $P$($\chi^2_{SF}$) of yielding the observed $\chi^2$ in
the fit to the optical-nearIR bands.  The values with
$P$($\chi^2_{SF}$)>5\% are shown for each object in
Fig.~\ref{fig:limits} in blue shades, while in grey are shown those
with lower probability.  Clearly, for each object the SED fitting SFRs
are spread over a large range of potential values, so that clear-cut
conclusions are difficult to reach.  In addition, the values inferred
from the far-IR models span almost a decade. For the sake of clarity,
we show the full range of far-IR models only for the stacks, shown in
Fig.~\ref{fig:limits_stacks}, but the same uncertainty applies to each
individual source.  Given the uncertainty associated with the choice
of the IR template and the arbitrariness of the 5\% threshold chosen
to accept a solution as "plausible", it is almost impossible to
extract a well-defined statistics.  However, it is clear from the
comparison reported in Fig.~\ref{fig:limits} that ALMA-based SFRs lend
crucial support to the passive nature of our candidates.

Given these uncertainties, we apply here conservative criteria to
confirm our candidates on an individual basis. We discard the S18 FIR
model as, on the basis of the motivations above, it is likely
inappropriate to describe our sources. We adopt our reference model
M10, but we note that the results are unchanged with the assumption of
the P08 FIR template, which predicts the highest SFR (with the
exception of S18).  We assume 3$\sigma$ submillimetric limits for all
sources.  We apply a criterion based on the requirement that all the
star--forming solutions with $P$($\chi^2_{SF}$)>5\% are above the ALMA
predictions.  We find that in 9 out of the total sample of 26
candidates (i.e., 35\%), 5 of which belonging to the most secure $S_1$
subsample (i.e., 63\%), the star--forming fits predict star formation
rates that are above, and often significantly larger than, the
estimates derived from ALMA, implying that such solutions are
implausible.  We mark these individually confirmed candidates with a
'$C$' in Table~\ref{tab:list}. We note that additional two sources
(ID23626 and ID5592) just do not pass the selection, but the bulk of
their optical fit solutions are anyhow above the 3$\sigma$ curve. We
also note that even the adoption of the most extreme, though likely
inappropriate, model of S18 with $R_{SB}$=$4$, results in the
exclusion of only two of the 9 selected sources (not belonging to
$S_1$).  The fact that most SED solutions are so high is explained by
the very red SEDs of these objects, which demand large amounts of dust
to be fitted with a star--forming template.  For this reason, this
test is conservative itself as the SFRs obtained from the optical fits
are known to be likely underestimated for extremely red and dusty
sources \citep[e.g.,][]{santini09}.  For the remaining, unconfirmed
candidates the comparison is inconclusive, as the values of SFR
predicted from the fit are lower than for the other sources, often
because of the lower amount of dust necessary to fit the observed
colour, and cannot be excluded by ALMA limits.  We only find a couple
of exceptions (ID7688 and ID15457) where ALMA-based SFRs are formally
consistent with the star-forming solutions of the optical fit.

\subsection{Validation of the whole population in a statistical sense}

While only 9 candidates can be individually confirmed with high
confidence, more information can be drawn from the data that can be
used to extract a statistical evaluation of the sample.  To this aim,
we adopt 1$\sigma$ limits. We notice that, according to a Gaussian
statistics, in 16\% of the undetected sources the limit may be too
optimistic, and these sources may be erroneously classified as
passive. However, this would not change the global results.

\begin{figure*}
	\includegraphics[angle=270,width=2\columnwidth,clip,viewport=0 0 320 830]{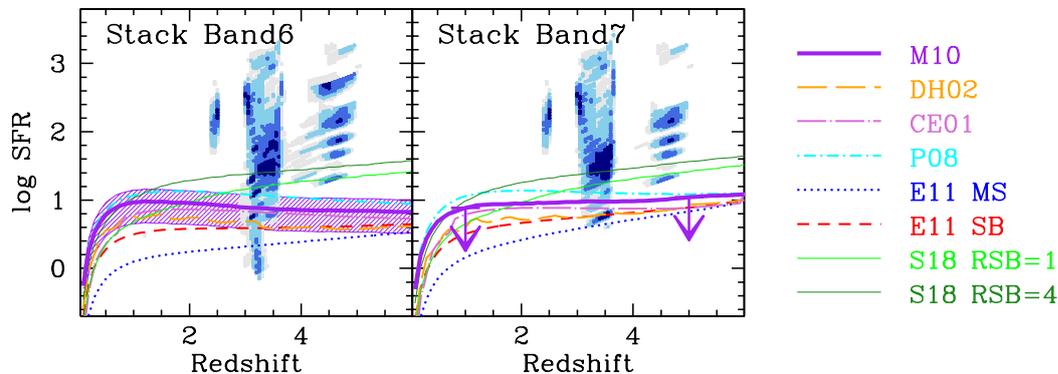}
        \caption{ SFR confidence interval ({\it left}) and 1$\sigma$
          upper limits on the SFR ({\it right}) at different redshifts
          given the ALMA observations, according to the predictions of
          the different templates adopted, for stacked sources in Band
          6 ({\it left}) and Band 7 ({\it right}) at all
          redshifts. Our reference model (M10) is shown by thick
          purple lines, while the others are represented by thin lines
          colour-coded and characterized by different linestyles
          according to the legend. The solutions of the optical fit
          (colour-coded as in Fig.~\ref{fig:limits}) have been plotted
          for all stacked sources.  }
    \label{fig:limits_stacks}
\end{figure*}

First of all, we take advantage from the stacking results to study the
population as a whole. We perform a similar test on the stacked
sources in the two ALMA bands. The stacked SFRs are compared with the
collection of star-forming solutions of all sources included in the
stacks.  The results are shown in Fig.~\ref{fig:limits_stacks}.  We
can claim that our candidates are on average consistent with being
passive, i.e. the ALMA (limiting) SFR is lower than the star--forming
solutions of the optical fit (we note that the low SFR tail in the
solutions are essentially given by one single source in each of the
stacks, i.e., ID10759 for Band 6 and ID2608 for Band 7).

A second, somewhat independent approach to statistically validate the
sample is the comparison of the ALMA derived SFRs with those predicted
by placing the objects on the star--forming MS.  Rather than relying
on the SFR derived from the fit to the optical-nearIR bands, in this
case we use the stellar mass estimated for our candidates (that is
usually considered a more robust measurement than the star formation
rate, \citealt{santini15}) and evaluate whether these objects have a
SFR lower than their siblings of the same stellar mass - i.e., if they
lie on the observed MS for star--forming galaxies, or below it. The
result of this comparison is shown in Fig.~\ref{fig:ms}, where we plot
the observed (i.e., not corrected for the Eddington bias) MS of
star--forming galaxies at the same redshifts as inferred from the HST
Frontier Fields data by \cite{santini17}.  Of the six marginally
detected sources, three fall in the quiescent area, i.e. are located
below the lower $3\sigma$ percentile of the distribution of
star-forming galaxies, one is 1$\sigma$ below the MS and two (ID7688
and ID15457, none of the two belonging to the most secure $S_1$
sub-sample) are consistent with the MS. However, the latter have with
huge error bars, especially extending into the passive region of the
diagram, that prevent any conclusion. Three/nine among the upper
limits indicate that the candidates are at least 3$\sigma$/1$\sigma$
below the MS. For only  a couple of sources the limit on the SFR
falls significantly above the MS, basically because of  their low
stellar mass. In total, given their stellar mass, the SFRs predicted
by ALMA place  13 (6) candidates, 6 (4) of which belonging to the
$S_1$ sample, at least 1$\sigma$ (3$\sigma$) below the MS  or
  around this threshold.  The passive classification is on average
confirmed at 1$\sigma$ for at least  50\% of the sample.  We note
that, with the exception of the two extreme templates of S18 (see
discussion above), these numbers are solid against the uncertainty in
the modelling of the FIR spectrum.

On the upper panels of Fig.~\ref{fig:ms}, we also show the average SFR
derived from stacking all undetected sources observed in the same ALMA
band (whose values are listed in the lower part of
Table~\ref{tab:sfr}) as large red symbols.  The average sources
observed in both bands lie below the $2\sigma$ lower envelope of the
distribution  and in one case below $3\sigma$, suggesting that
our candidates undetected by ALMA are on average correctly classified
as passive even when submillimeter data are not deep enough to draw
conclusions.

\subsection{Final considerations}

\begin{figure} 
    \includegraphics[angle=0,width=\columnwidth,clip,viewport=30 -20 680 450]{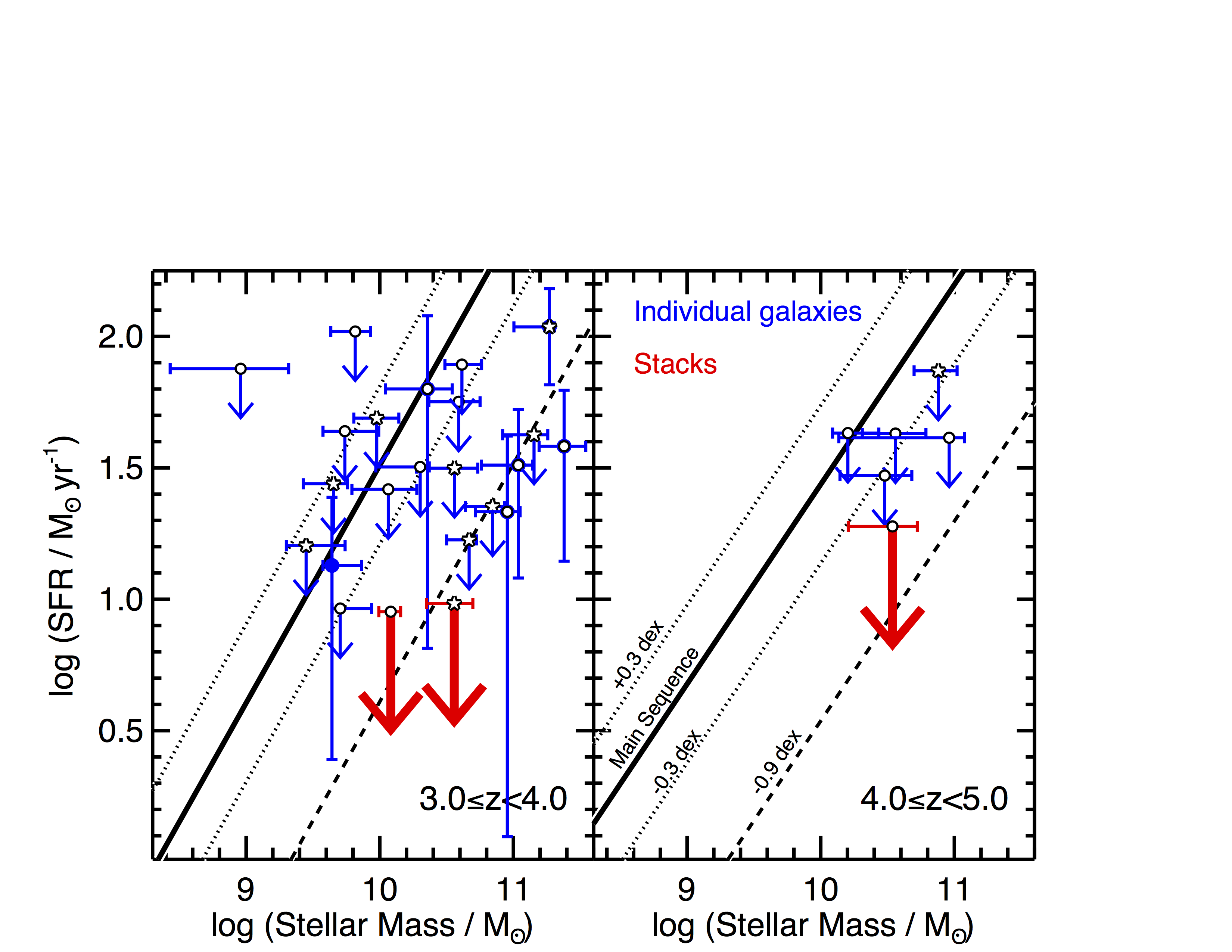}
    \includegraphics[angle=0,width=\columnwidth,clip,viewport=30 -20 680 450]{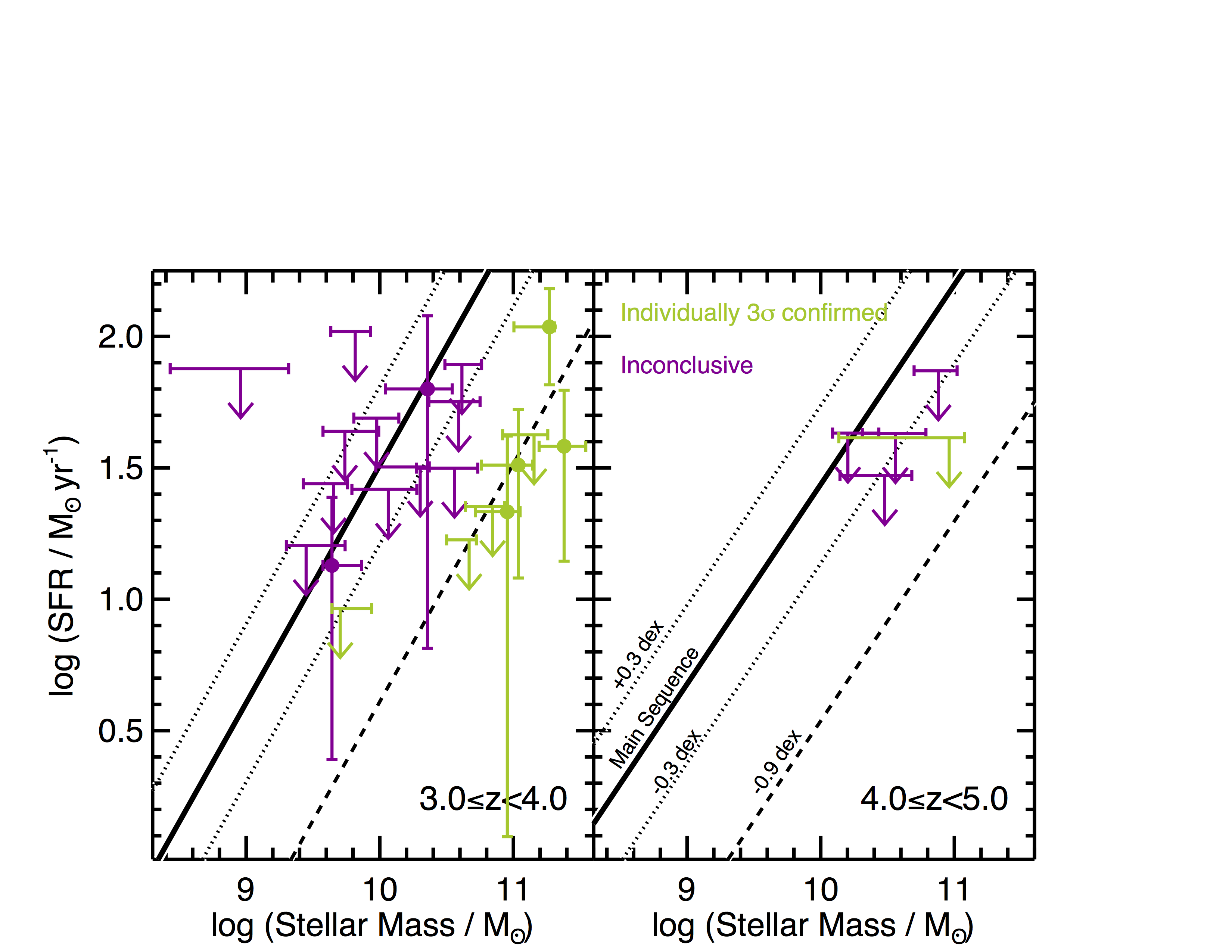}
    \caption{Location of the passive candidates on the SFR-stellar
      mass diagram, in two redshift bins, based on their ALMA
      SFR. Arrows represent 1$\sigma$ upper limits on the SFR. Solid
      lines show the observed MS (i.e., not corrected for the
      Eddington bias) inferred from HST Frontier Field data by
      \protect\cite{santini17}.  Dotted lines are 1$\sigma$ above and
      below the MS (estimated from the observed 0.3 dex scatter),
      while the dashed line is 3$\sigma$ below it. In the {\it upper}
      panels, white circles and stars denote observations in Band 6
      and Band 7, respectively. Blue symbols show the individual
      sources and large and thick red symbols show the stacks of the
      undetected sources in each redshift bin. In the {\it lower}
      panels, green symbols represent objects whose passive nature is
      individually and robustly confirmed by ALMA at
      $\geq$3$\sigma$, i.e., objects whose SFR predicted by ALMA at
      any redshift is below any possible star--forming solutions of
      the optical fit with an acceptable probability
      ($P$($\chi^2_{SF}$)>5\%).  Purple symbols show objects for which
      ALMA observations may be consistent with the star--forming
      solutions with $P$($\chi^2_{SF}$)>5\%, but that are inconclusive
      because of too shallow sub-mm data. See text and
      Fig.~\ref{fig:limits} for details.  }
\label{fig:ms}
\end{figure}

It is important to remark that the two analyses yield consistent
results, in that they both identify a subsample of more secure passive
candidates and one made of sources for which the inferred limits on
the SFR are not stringent enough to draw firm conclusions, although
the stacking results seem to corroborate their passive nature. This
can be clearly seen on the lower panels of Fig.~\ref{fig:ms}: the
objects for which the submm-based SFRs are (much) lower than those
allowed by the star--forming solutions of the optical fit (green
symbols) populate the passive region of the SFR--stellar mass diagram
while those whose ALMA limits do not exclude the star--forming
solutions (purple symbols) lie around (or above) the MS.  As
mentioned, with both approaches, the results of the stacks yield
tighter constraints.

It is interesting to note that the present analysis corroborates the
interpretation of galaxies recently quenched by the emission of their
still actively radiating nucleus for two of our candidates. Indeed, to
exclude contamination from red, dusty sources, in M18 we searched for
FIR emission on {\it Herschel} observations, and found a detection for
two of the strongest candidates (ID10578 and ID3973). After a careful
analysis of their optical and X-ray emission, we attributed the FIR
emission to a dust-obscured AGN hosted at their centre. The much
fainter and marginal detection at submm wavelengths confirms that {\it
  Herschel} fluxes are likely caused by host dust heated by the
central nucleus rather than cold dust heated by newly formed stars.

\section{Summary and conclusions} \label{sec:concl}

In this paper we have presented a follow-up analysis of the passive
galaxy candidates selected at high redshift ($z$=3--5) by our previous
study (M18) by means of SED fitting of optical--nearIR bands. Starting
from a sample of 30 candidates in the GOODS-S field, we found ALMA
archival observations for 26 of them, in Bands 6 and Band 7.  None of
the individual sources is robustly (i.e., $>$3$\sigma$) detected in
the sub-mm, the large majority of sources being below the noise
level. Indeed, the few marginally detected sources are consistent with
belonging to the tail of the Gaussian noise distribution. No
significant detection is obtained even by stacking sources observed in
the same ALMA band. This allows us to exclude the possibility that a
significant fraction of undetected objects has actually flux at
>1$\sigma$. From the flux and rms measured on the ALMA images, we
derived estimates of, or in most cases upper limits on, the SFR, that
we use to validate the passive nature of our candidates, both
individually and in a statistical sense.

Firstly, we compared the sub-mm based SFRs with the star--forming
secondary solutions of the optical fits.  For nine candidates the
star--forming solutions are rejected by the ALMA observations adopting
the most conservative assumptions, i.e., adopting 3$\sigma$ upper
limits and the FIR models providing the highest SFRs (with the
exception of S18 model that we deem inappropriate to describe our
sources). These sources are individually confirmed with high
confidence.  Secondly, we used the ALMA-based (limits on the) SFRs to
compare the location of our candidates with respect to the star
formation MS, given their stellar mass: 50\% of the candidates are
placed below the 1$\sigma$ distribution of star--forming galaxies, and
23\% (6 out of 26) fall in the quiescent area (i.e., 3$\sigma$ below
the MS).  The results of the two tests overlap very nicely.  While for
the remaining, unconfirmed, candidates the comparison is inconclusive
because the available sub-mm data is too shallow to draw firm
conclusion, the stacking results suggest an overall passive nature for
our sample.

Although the exact quantification of the fraction of confirmed
candidates depends on the details of the analysis and of the models,
we can reach the following conclusions from our study:
\begin{itemize}
\item ALMA observations lend decisive evidence to the quiescent nature
  of our passive candidates, that clearly show a distribution of SFR
  that is inconsistent with the typical one at these redshifts;
\item currently available ALMA archive observations are not deep
  enough to individually confirm most of our candidates with high
  confidence; however:
\item we can individually confirm 9 candidates out of 26 (35\%)
  adopting conservative assumptions;
\item the stacking analysis and the lack of reliable detections
  corroborate the passive nature of the remaining part of the sample,
  at least in a statistical sense;
\item at least half of the sample is located at least 1$\sigma$ below
  the Main Sequence;
\item these results confirm the existence of passive galaxies in the
  early Universe ($z>3$)

  and
\item validate the robustness and reliability of the selection
  technique developed by our previous analysis (M18), in particular
  when the most conservative selection criteria are adopted.
\end{itemize}

In the next future, JWST observations not only will improve the
selection of passive galaxy candidates at high $z$ (M18), but will
also make it possible to finally confirm them by means of a
spectroscopic analysis.

\section*{Acknowledgements}
We thank the anonymous referee for his/her helpful comments, and we 
thank M. Dickinson and M. Ginolfi for useful discussion regarding
ALMA data reduction.  The research leading to these results has
received funding from the European Union Seventh Framework Programme
ASTRODEEP (FP7/2007-2013) under grant agreement No. 312725. This paper
makes use of the following ALMA data: ADS/JAO.ALMA\#2012.1.00173.S,
ADS/JAO.ALMA\#2012.1.00869.S, ADS/JAO.ALMA\#2013.1.00718.S,
ADS/JAO.ALMA\#2013.1.01292.S, ADS/JAO.ALMA\#2015.1.00098.S,
ADS/JAO.ALMA\#2015.1.00543.S, ADS/JAO.ALMA\#2015.1.00664.S,
ADS/JAO.ALMA\#2015.1.00870.S, ADS/JAO.ALMA\#2015.1.01074.S,
ADS/JAO.ALMA\#2015.1.01495.S.  ALMA is a partnership of ESO
(representing its member states), NSF (USA) and NINS (Japan), together
with NRC (Canada), NSC and ASIAA (Taiwan) and KASI (Republic of
Korea), in cooperation with the Republic of Chile. The Joint ALMA
Observatory is operated by ESO, AUI/NRAO and NAOJ.


\bibliographystyle{mnras}

   \bsp	
\label{lastpage}
\end{document}